\documentclass[runningheads]{llncs}
\usepackage[T1]{fontenc}
\usepackage{graphicx}

\usepackage{booktabs}
\begin{document}
\title{PUUMA (Placental patch and whole-Uterus dual-branch U-Mamba-based Architecture): Functional MRI Prediction of Gestational Age at Birth and Preterm Risk}

\titlerunning{PUUMA: Functional MRI Prediction of GA at Birth and Preterm Risk}
%
%
\author{Diego Fajardo-Rojas\inst{1,2,6} \and
Levente Baljer\inst{3} \and
Jordina Aviles Verdera\inst{4}  \and
Megan Hall \inst{1,5}  \and
Daniel Cromb \inst{1}  \and
Mary A. Rutherford \inst{1} \and
Lisa Story \inst{1,5}  \and
Emma C. Robinson \inst{2}  \and
Jana Hutter \inst{1,5} 
}
\authorrunning{D. Fajardo-Rojas et al.}
%
\institute{Early Life Imaging department, King's College London, London, UK \and
Biomedical Computing department, King's College London, London, UK
\and Neuroimaging department, King's College London, London, UK
\and Smart Imaging Lab, Radiological Institute, University Hospital Erlangen, Erlangen, Germany
\and Women's Health department, School of Life Course and Population Sciences, King's College London, London, UK
\and
\email{diego.fajardo\_rojas@kcl.ac.uk}\\
}

\maketitle              

\begin{abstract}
Preterm birth is a major cause of mortality and lifelong morbidity in childhood. Its complex and multifactorial origins limit the effectiveness of current clinical predictors and impede optimal care. In this study, a dual-branch deep learning architecture (PUUMA) was developed to predict gestational age (GA) at birth using T2* fetal MRI data from 295 pregnancies, encompassing a heterogeneous and imbalanced population. The model integrates both global whole-uterus and local placental features. Its performance was benchmarked against linear regression using cervical length measurements obtained by experienced clinicians from anatomical MRI and other Deep Learning architectures. The GA at birth predictions were assessed using mean absolute error. Accuracy, sensitivity, and specificity were used to assess preterm classification. Both the fully automated MRI-based pipeline and the cervical length regression achieved comparable mean absolute errors (3 weeks) and good sensitivity (0.67) for detecting preterm birth, despite pronounced class imbalance in the dataset. These results provide a proof of concept for automated prediction of GA at birth from functional MRI, and underscore the value of whole-uterus functional imaging in identifying at-risk pregnancies. Additionally, we demonstrate that manual, high-definition cervical length measurements derived from MRI —not currently routine in clinical practice— offer valuable predictive information. Future work will focus on expanding the cohort size and incorporating additional organ-specific imaging to improve generalisability and predictive performance.

\end{abstract}
\section{Introduction}

Preterm birth is defined as a live birth before 37 completed weeks of gestation. The global annual incidence is estimated at 9.9\% or 13.4 million \cite{ohuma}. This makes prematurity the leading cause of death among children under five, corresponding to 17.7\% of the 5.3 million annual deaths in this age group. Complications arising from preterm birth are also the primary cause of neonatal mortality, accounting for 36\% of these deaths \cite{perin}. 

Preterm birth is classified into three subcategories: extremely preterm (EPT, less than 28 weeks), very preterm (VPT, 28 to 32 weeks), and late preterm (LPT, 32 to 37 weeks). This categorisation is relevant since the risk of mortality among preterm babies is inversely related to gestational age (GA) at birth, decreasing from over 82\% for babies born at 22 weeks to less than 5\% for those born at 29 weeks or later \cite{ancel,santhakumaran}. The same inverse relation holds for the incidence and severity of short- and long-term consequences, which include a heightened risk of acute complications (such as infection, respiratory, and neurological disorders) and chronic neurodevelopmental, sensory, and cognitive impairments \cite{Costeloe2012ShortTO,vanes,allen_outcomes}.

Another categorisation of preterm birth is defined by its clinical presentation: medically induced (iatrogenic) or spontaneous \cite{moutquin}. Indicators for iatrogenic preterm birth are well characterised and include fetal growth restriction and pre-eclampisa. In contrast, spontaneous preterm birth is considered to be a multifactorial syndrome \cite{goldenberg} with varied, complex aetiologies that are poorly understood \cite{frey}. Causes include infection or inflammation, vascular disease, cervical injury, and uterine overdistention \cite{goldenberg,suff_csections}.

The placenta plays a key role in the mechanisms leading to preterm birth, particularly in the context of infection and inflammation. Microbial invasion can reach the placenta and fetal membranes, triggering inflammatory responses that weaken tissues and can precipitate early labour \cite{habelrih}. This is especially relevant in cases of preterm premature rupture of membranes (pPROM), where the disruption of membrane integrity further increases susceptibility to infection and inflammation \cite{cervantes}. Furthermore, conditions such as chorioamnionitis are found in a large proportion of extremely preterm deliveries, underscoring the prevalence of this relationship \cite{kemp}. Beyond infection, placental vascular and immune functions also shape the in utero environment, further influencing the timing of birth \cite{kemp,cervantes}.

Current clinical prediction tools—including previous history of preterm birth, cervical length obtained by trans-vaginal ultrasound, and fetal fibronectin testing—remain limited: they may not apply to all pregnancies, rely on invasive procedures, and are operator-dependent or unavailable \cite{suff,goldenberg,hologic}. Fetal MRI, in contrast, offers safe, non-invasive, high-resolution assessment of placental structure and function, with T2* relaxometry providing insight into oxygenation and tissue microstructure \cite{hutter,story_antenatal}.

Given the multifactorial and poorly understood nature of preterm birth, few studies have attempted direct prediction of GA at birth as a continuous outcome. Instead, most prior work has focused on classifying preterm birth risk by applying machine learning models to specific data sources, such as uterine electromyography and transvaginal ultrasound cervical length measurements \cite{prema,wlodarczyk_unet}.

Accurate predictions of GA at the time of birth could improve the care of pregnant women and their babies, specifically for preterm cases. Such improvements range from ensuring women are transferred to appropriate neonatal care facilities — which would lead to decreased neonatal mortality and care costs \cite{story_antenatal} — to the effective implementation of therapies that mitigate the effects of prematurity. Concretely, cortiscosteroid administration promotes lung maturity and helps reduce intraventricular haemorrhage, but its timing is crucial. Results of this therapy are optimal within a week before delivery, and repeated doses increase the risk of adverse effects including reduction in birthweight \cite{story_impactuk}.


\noindent \textbf{\textit{Contributions: }} In this study PUUMA, a modified version of the U-Mamba \cite{U-Mamba} segmentation architecture, is implemented to perform a regression task for predicting GA at birth from T2* whole-uterus images. The approach leverages the role of the placenta in the aetiology of preterm birth by using high-definition sub-volumes of this organ. Additionally, the predictive value of cervical length measurements obtained from anatomical MR images by an experienced clinician is assessed. Results demonstrate that both fully automated prediction and simple modeling based on manual cervical length measurements are valuable approaches toward the accurate prediction of preterm birth. Both approaches underscore the value of MRI, which enables analysis on detailed images, as well as reliable cervical length measurements not available with standard clinical methods.

\section{Methods}

\subsection{Data Acquisition}

The data used in this study were collected as part of four ethically approved research projects: [14/LO/1169, Placenta Imaging Project, Fulham Research Ethics Committee, approval received September 23, 2016], [19-SS-0032, Inflammation study in pregnancy, South East Scotland Ethics Committee, approval  received  March  7,  2019], [22/YH/0210, NANO: MRI aNd Antenatal luNg develOpment, South Yorkshire Research Ethics Committee, approval received 4 Oct 2022], and [21/SS/0082, Individualised Risk prediction of adverse neonatal outcome in pregnancies that deliver preterm using advanced MRI techniques and machine learning, South East Scotland Ethics Committee, approval received March 2022]. Informed consent was secured from all participants.

MR imaging was performed on a 3T Philips Achieva scanner between 15 and 40 weeks of gestation, using a 32-channel coil and protocols consistent across studies. Each scan included anatomical T2-weighted images and functional sequences; for this work, only T2* relaxometry data were analysed. Despite standardised acquisition, pooling from multiple studies yielded a heterogeneous dataset.

Anatomical images were acquired with a 2D Turbo-Spin-Echo sequence, followed by B0 field mapping for shimming, and whole-uterus functional imaging with multi-echo gradient-echo EPI. No motion correction was required, as all echoes per slice were acquired within 200 ms. Quantitative T2* maps were generated via mono-exponential fitting and clipped at 300 ms. Placental segmentation masks were automatically generated using a nnU-Net \cite{nnunet}, trained on expert manual annotations \cite{hall}.

Patient demographics, medical and obstetric history, and outcomes—including GA at birth, birth weight centile, and major complications—were recorded. Cervical length measurements were manually extracted by an experienced clinician from the anatomical T2-weighted images, though this measurement was not systematically performed for all participants.

\begin{figure}
\centering
\includegraphics[width=0.75\textwidth]{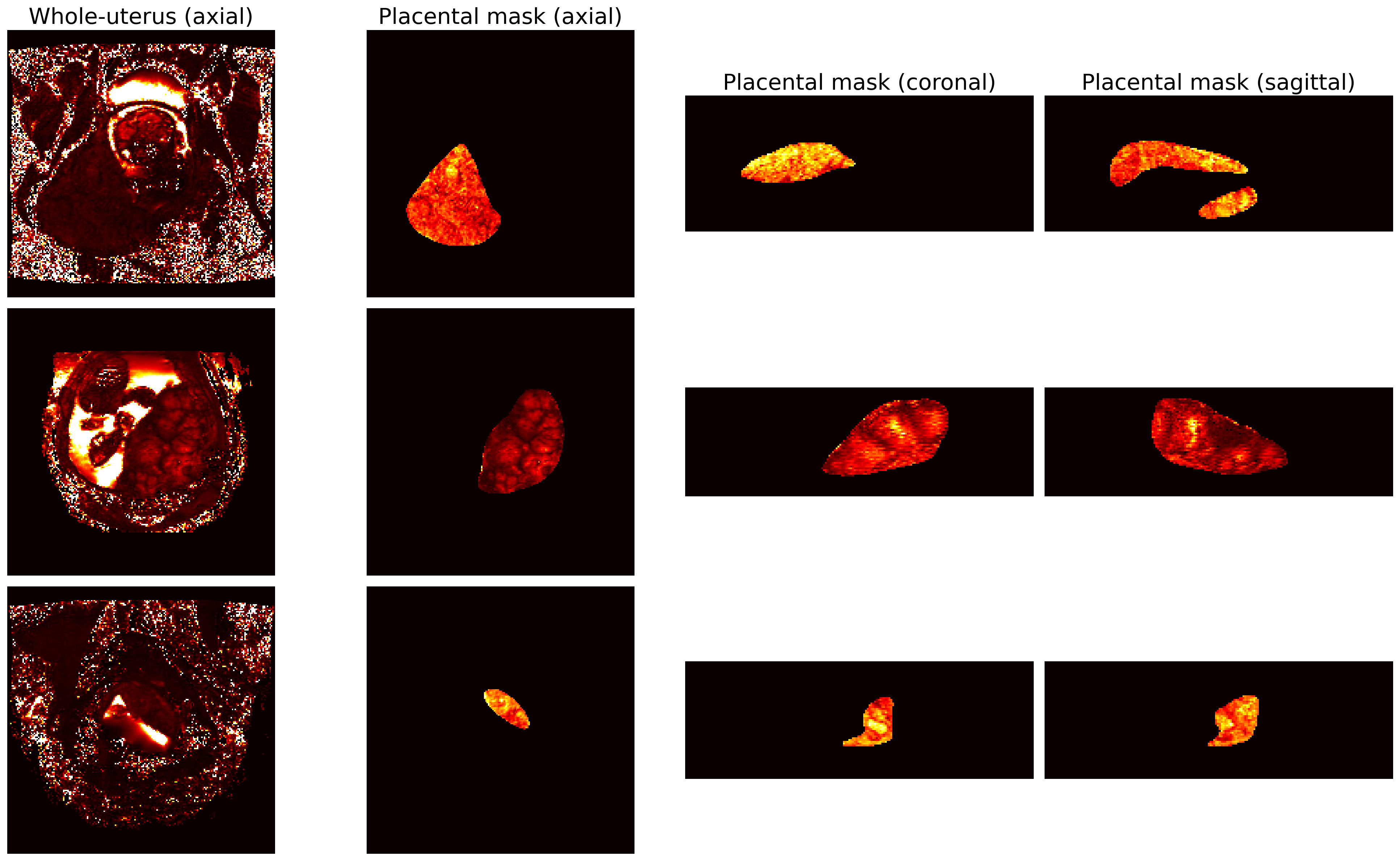}
\caption{Example T2* MR images from the dataset, illustrating the heterogeneity of whole-uterus volumes (left column) and corresponding masked placentas (axial, coronal, and sagittal views; center and right columns). Substantial variation is observed in placental size, shape, and positioning across subjects, reflecting the diversity and complexity of the study dataset.} 
\label{fig:data_overview}
\end{figure}

\subsection{Data Preprocessing}

All fetal scans acquired after 37 weeks GA (term patients) were removed from the dataset, as well as those missing a record of GA at delivery. Cases with artifacts or inaccurate segmentations were excluded after visual inspection of the T2* volumes and their respective placenta masks. After this selection process, a total of $n=295$ cases remained. Each case included a record of GA at both scan and birth, a whole-uterus T2* volume with a corresponding placenta mask, and, for most cases, a manually obtained cervical length measurement from an anatomical T2-weighted image. Figure~\ref{fig:data_overview} shows examples of whole-uterus volumes and their corresponding masked placentas.


\begin{figure}
\centering
\includegraphics[width=0.8\textwidth]{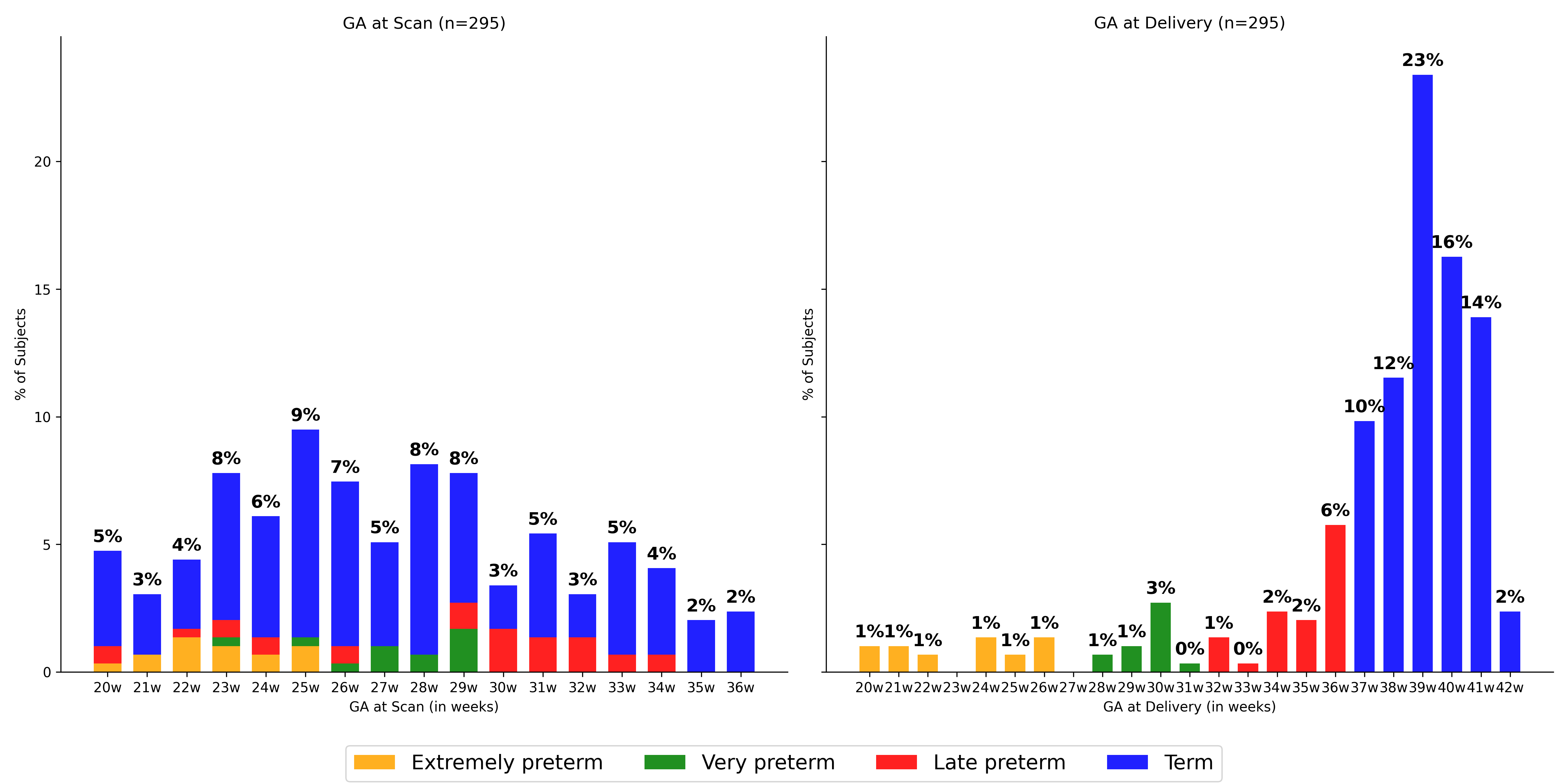}
\caption{Distribution of gestational age (GA) at scan (left) and at delivery (right) for all cases included in the study ($n=295$). The majority of subjects delivered at term, with much smaller proportions in the extremely, very, and late preterm categories, highlighting the pronounced class imbalance present in the dataset. This imbalance poses significant challenges for preterm birth and GA at birth prediction.} 
\label{fig:ga_histograms}
\end{figure}

As illustrated in Figure~\ref{fig:ga_histograms}, the dataset is highly imbalanced, with a much greater proportion of term cases compared to extremely, very, and late preterm births. To mitigate the impact of this imbalance during model development, the dataset was split into training ($n_{train}$=243), validation ($n_{val}$=26), and test ($n_{test}$=26) sets using an 8:1:1 ratio, ensuring that each set contained an equal proportion of extremely preterm, very preterm, late preterm, and term cases. All subjects in the test set had an associated cervical length measurement.


\subsection{Regression of GA at birth}

Introduced in the context of sequence modeling, the Mamba architecture \cite{gu2024mambalineartimesequencemodeling} employs State Space Sequence Models \cite{gu2022efficientlymodelinglongsequences} to capture long-range dependencies while maintaining linear computational complexity. U-Mamba incorporates Mamba blocks into the U-Net framework, combining the local feature extraction of CNNs with enhanced global context modeling. This architecture has been shown to outperform state-of-the-art CNN and Transformer-based networks, particularly on volumetric CT and MR segmentation tasks \cite{U-Mamba}.

\subsubsection{Inputs.} The main objective of this work is to predict GA at birth from the shape and function of the placenta, inferred from T2* whole-uterus volumes. To achieve this, the proposed architecture takes as input both downsampled whole-uterus volumes (size $128\times128\times64$) and high-resolution patches (size $16\times16\times16$), that get randomly sampled from locations that overlap with 33\% of the placental mask. By combining these inputs, the model is designed to integrate local placental features with global placental characteristics—such as topology and whole-organ shape—as well as broader uterine functional information. 

\subsubsection{Network Architecture.}

PUUMA (Placental patch and whole-Uterus dual-branch U-Mamba-based Architecture), a bespoke architecture featuring two complementary global and local branches for prediction of GA at birth, was implemented. The global branch, closely resembling the original U-Mamba framework, processes the resized whole-uterus images through a symmetric encoder-decoder pathway to output a prediction of the placental segmentation mask. A prediction head (fully connected layer) is then added to the bottleneck of the network to both regress GA at birth and classify individuals into preterm birth categories (EPT,VPT,LPT, T).

\begin{figure}[htbp]
\centering
\includegraphics[width=0.75\textwidth]{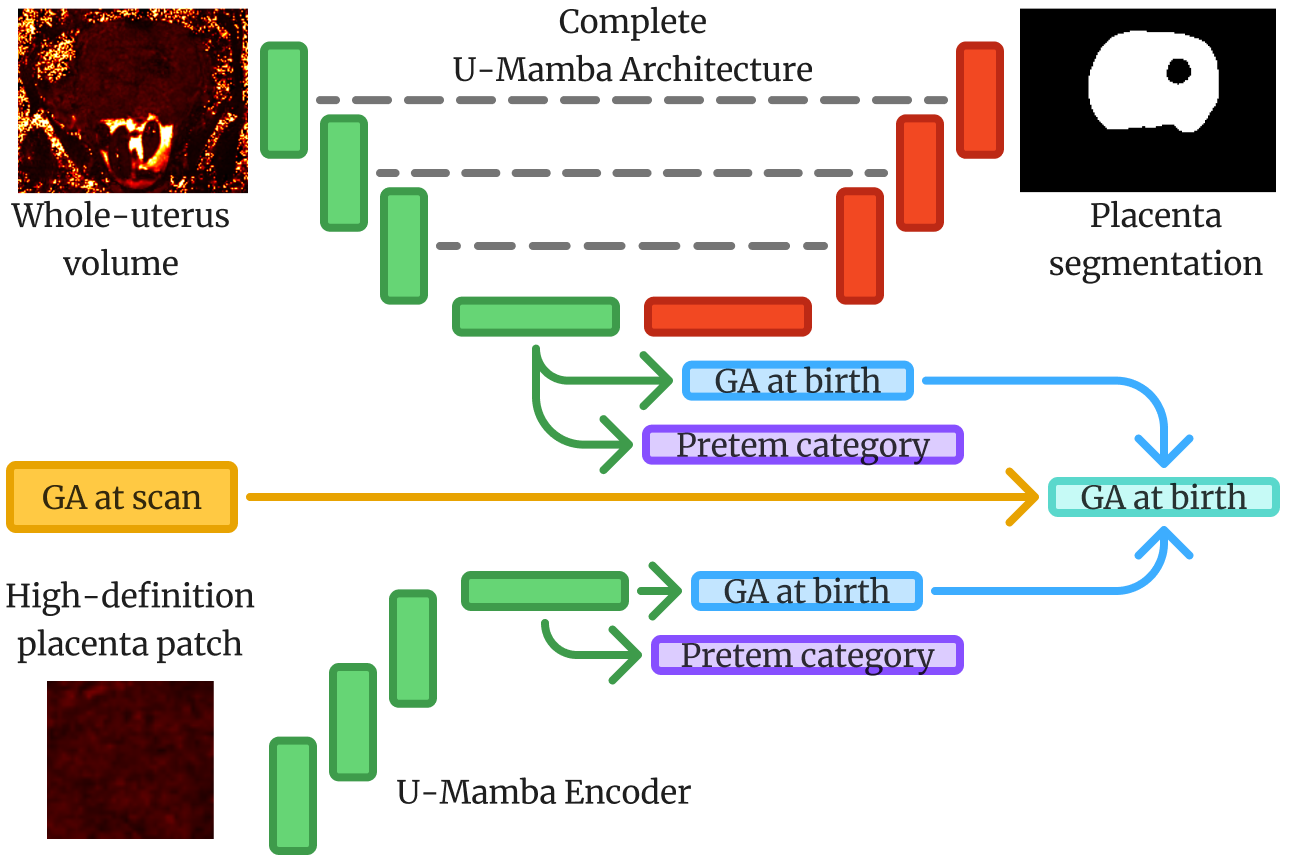}
\caption{Overview of the PUUMA (Placental patch and whole-Uterus dual-branch U-Mamba-based Architecture) design for predicting gestational age (GA) at birth. The global branch (top) processes the whole-uterus T2* MRI volume using the U-Mamba encoder-decoder architecture to extract global features and predict placental segmentation, as well as to perform regression and classification at the bottleneck. The local branch (bottom) extracts high-resolution patches of the placenta and processes them with a U-Mamba encoder to capture detailed local features for regression and classification. Predictions from both branches, together with gestational age at scan, are combined to yield the final prediction.} 
\label{fig:net-architecture}
\end{figure}

In parallel, the local branch focuses on high-resolution placental patches, making use of only the U-Mamba encoder to directly regress GA at birth and classify samples into preterm categories from the latent patch representation. Predictions from both branches, along with gestational age at scan, are then concatenated and a fully connected layer is used to yield the final prediction. This architectural design enables the model to integrate detailed local placental features with global placental characteristics—such as topology and whole-organ shape—as well as broader uterine functional information. Figure~\ref{fig:net-architecture} illustrates the network design. 


\subsubsection{Training.}

To increase data diversity and improve generalization, data augmentation was applied to whole-uterus volumes before extracting patches, including applying affine and elastic transformations, zoom, and contrast adjustment, as well as simulating bias fields and Gaussian noise. Each augmentation was applied independently and probabilistically. To address class imbalance, samples from each preterm birth temporal category were drawn with an inversely proportional frequency during training.

 The network was trained with a batch size of 1 using the Adam optimiser with an initial learning rate of $0.0001$, annealed with a learning rate scheduler. The loss function was a weighted sum of mean squared error (for GA at birth regression), Dice and binary cross-entropy losses (for placenta segmentation, following \cite{U-Mamba}), and categorical cross-entropy (for preterm classification). The segmentation and classification tasks serve as auxiliary objectives to help guide the network's feature learning, with the main task being regression of GA at birth.

Model selection and validation were performed using the validation set. For inference, a Sliding Window approach was used to sample multiple placental patches from each whole-uterus image, generating pairs (whole-uterus image, patch) as input to the network. Only those pairs in which the patch contained more than $33\%$ placental tissue were included in the final prediction. The network’s output for GA at birth was calculated as the mean prediction across all valid pairs . The best-performing model was selected based on the lowest regression loss (mean squared error) observed on the validation set, with checkpoints saved every 50 batches and then every 5 batches during fine-tuning.

\subsection{Evaluation.}

\subsubsection{Baselines:}

Of the $243$ cases in the training set, $n_{train_{cl}}=170$ had an available cervical length measurement. A linear regression to predict GA at birth, based exclusively on cervical length values, was fitted using this set. Network performance was benchmarked both against the PUUMA global branch —consisting primarily of the original U-Mamba architecture with added regression and classification heads— and a simple U-Net, similarly modified to include both regression and classification heads at the bottleneck layer, and matching the architectural hyperparameters of the global branch (see Table \ref{tab:net_hyperparams}). Both models took whole-uterus T2* volumes as input.

\begin{table}[ht]
    \centering
    \caption{Model Hyperparameters.
    L. = Local Branch, G. = Global Branch, F.C.L. = Fully Connected Layer, Dim. = Dimension, Params. = Trainable Parameters.}
    \label{tab:net_hyperparams}
    \begin{tabular}{|l|c|c|c|c|c|c|}
        \hline
        {\scriptsize \textbf{Model}} & 
        {\scriptsize \textbf{L. Depth}} & 
        {\scriptsize \textbf{G. Depth}} & 
        {\scriptsize \textbf{L. Latent Dim}.} & 
        {\scriptsize \textbf{G. Latent Dim.}} & 
        {\scriptsize \textbf{F.C.L. Dim.}} & 
        {\scriptsize \textbf{Params.}} \\
        \hline
        PUUMA & 3 & 6 & 4096 & 5120 & $5120 \times 16$ & 8.5M \\
        \hline
        U-Mamba & N/A & 6 & N/A & 5120 & $5120 \times 16$ & 8M \\
        \hline
        U-Net & N/A & 6 & N/A & 5120 & $5120 \times 16$ & 4.8M \\
        \hline
    \end{tabular}
\end{table}

\subsubsection{Metrics:} Performance of all models was assessed using mean absolute error (MAE) measured in gestational weeks. 
Additionally, cases were classified as term ($\geq 37$ weeks) or preterm ($<37$ weeks), and accuracy, sensitivity, and specificity were reported.

\section{Results and Discussion}

Table~\ref{tab:comparison} presents the performance of all models on the test set. Since the central aim of this work is to identify pregnancies at risk of preterm birth, sensitivity is a particularly important metric. In terms of mean absolute error, the three leading approaches — PUUMA, linear regression on cervical length, and whole-uterus U-Mamba — achieved comparable results ($3.05 \pm 3.1$ weeks, $2.94 \pm 2.59$ weeks, and $2.95 \pm 3.65$ weeks). However, only PUUMA and the cervical length-based model attained the highest sensitivity for preterm detection (0.67), whereas the U-Mamba global branch alone missed more than half of the preterm cases. The baseline U-Net performed less favourably than PUUMA and cervical length-based regression in terms of MAE and sensitivity.

\begin{table}[ht]
    \centering
    \caption{Performance Comparison of Models}
    \label{tab:comparison}
    \begin{tabular}{|l|c|c|c|c|}
        \hline
        \textbf{Model} & \textbf{MAE (weeks)} & \textbf{Accuracy} & \textbf{Sensitivity} & \textbf{Specificity} \\
        \hline
        PUUMA & $3.05\pm3.1$ & $0.65$ & $\mathbf{0.67}$ & $0.65$ \\
        U-Mamba & $2.95\pm3.65$ & $0.73$ & $0.33$ & $\mathbf{0.85}$ \\
        LR on cervical length & $\mathbf{2.94\pm2.59}$ & $\mathbf{0.77}$ & $\mathbf{0.67}$ & $0.8$ \\
        U-Net & $3.98\pm3.6$ & $0.65$ & $0.5$ & $0.7$ \\
        \hline
    \end{tabular}
\end{table}

\begin{figure}[htbp]
\centering
\includegraphics[width=0.85\textwidth]{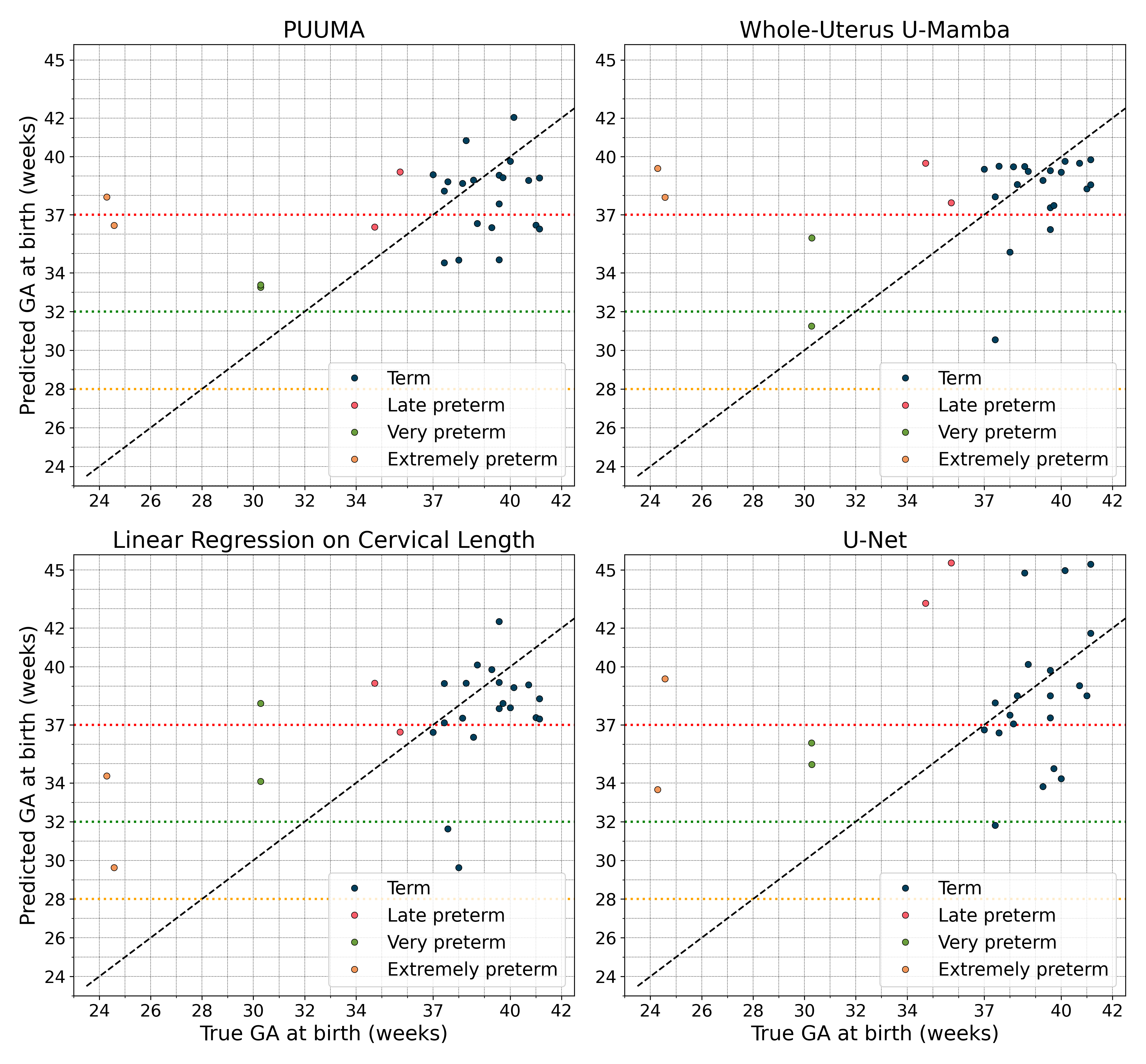}
\caption{Predicted versus true gestational age (GA) at birth for all test subjects, shown for each model. Points are colour-coded by preterm temporal category: term (blue), late preterm (red), very preterm (green), and extremely preterm (orange). Dashed black line indicates perfect prediction (identity). Coloured dotted lines delineate preterm category boundaries.} 
\label{fig:comparison}
\end{figure}

These results are visualised in Figure~\ref{fig:comparison}. Notably, PUUMA was the only model to predict both very preterm subjects as born before 34 weeks GA. Most models rarely classified extremely or very preterm cases correctly, tending to overestimate gestational age in these groups. The linear regression on cervical length produced consistent predictions for term births and performed comparatively better in both extremely preterm cases, uniquely classifying one as very preterm. This strong performance for term cases is consistent with the negative predictive value of cervical length reported in the literature \cite{suff}.

The classification performance of both PUUMA and the cervical length regression model is noteworthy given the heterogeneity and strong class imbalance in our dataset. For context, \cite{wlodarczyk_unet} developed an automated method for cervical length measurement from transvaginal ultrasound, but evaluated preterm birth prediction only on a well-balanced dataset using manual measurements instead of their automated ones. In contrast, both PUUMA and our cervical length-based model achieved strong results on a diverse and imbalanced cohort, underscoring their potential for practical value in real-world clinical settings.

\section{Conclusion and Future Work}

While the present study demonstrates the potential of MRI for preterm birth risk assessment—both through fully automated prediction using PUUMA and through cervical length measurements obtained by experienced clinicians—the small size of the test set limits the generalisability of these findings. Nevertheless, these results provide a proof of concept for leveraging functional and anatomical MRI in heterogeneous populations. Future work should focus on expanding the dataset size, as well as incorporating additional organ imaging, to improve model robustness and predictive accuracy.

\begin{credits}
\subsubsection{\ackname} This work was supported by funding from the EPSRC Centre for Doctoral Training in Smart Medical Imaging (EP/S022104/1) to Diego Fajardo-Rojas, from the Wellcome/EPSRC Centre for Medical Engineering (WT203148/Z/16/Z), a UKRI FLF (MR/T018119/1), and DFG Heisenberg funding through the High Tech Agenda Bavaria (502024488) to Jana Hutter, and from the NIHR Advanced Fellowship (NIHR3016640) and the MRC grant (MR/W019469/1) to Lisa Story. 

The authors acknowledge the invaluable help of the radiographers and midwives while acquiring the data presented here.

\subsubsection{\discintname}
The authors have no competing interests to declare that are relevant to the content of this article.
\end{credits}

%
%
%
\bibliographystyle{splncs04}

\bibliography{citation}

\end{document}